# Effect of temperature, pressure and aging time on the relaxation dynamics of $Bi_{0.9}Gd_{0.1}Fe_{0.9}Mn_{0.1}O_3$ system: Direct evidence of glassy state and pressure induced relaxor behavior

Satya N. Tripathy[1,2,3(a)], Zaneta Wojnarowska[1,3], Justyna Knapik[1,3], Arthur Chrobak[1],

Dillip K. Pradhan[4], Sebastian Pawlus[1,3] and Marian Paluch[1,3 (a)]

[1]*Institute of Physics, University of Silesia, Uniwersytecka 4, 40-007 Katowice, Poland.*

[2]*Department of Physics, Koneru Lakshmaiah Education Foundation, Deemed to be University, Hyderabad-500075, Telangana, India*
[3]*Silesian Center for Education and Interdisciplinary Research, 75 Pulku Piechoty 1A, 41-500 Chorzow, Poland.*
[4]*Department of Physics and Astronomy, National Institute of Technology, Rourkela-769008, India.*

## Abstract

The fundamental aspects of relaxation dynamics in $Bi_{0.9}Gd_{0.1}Fe_{0.9}Mn_{0.1}O_3$ multiferroic nano ceramic system have been reported. The study was carried out employing dielectric spectroscopy covering eight decades in frequency ($10^{-2}$ to$10^{6}$ Hz) and in a wide range of temperature (423 K to 153 K), pressure (0.1 MPa to 1765 MPa) and aging time (0 s to 80000 s). The temperature dependent dielectric response indicates three relaxations processes. Variable range hopping model of small polarons manifests the bulk conduction mechanism. The direct evidence of glassy feature is established below T ≈ 200 K by aging experiments. Our findings provide a potential connection between nearly constant loss features appearing below 200 K with fastest relaxation of magnitude 0.16 eV. We also discuss the time-temperature superposition behavior using modulus scaling. A pressure driven normal ferroelectric to relaxor behavior is witnessed above a critical pressure as a result of relative competition between short range and long range forces.

**Key Words:** Multiferroics, Relaxation dynamics, Glassy State, Nearly constant loss feature, Relaxor.

[(a)]Author of correspondence: satya.tripathy@smcebi.edu.pl and marian.paluch@us.edu.pl

## Introduction

The fundamental aspects of relaxation dynamics in the disorder solids include dipolar (localized) and/or electric conductivity (long-range/ delocalized) relaxation processes. The relaxation dynamics is governed by inter-molecular interaction, crystal symmetry, chemical composition, and external thermodynamic intensive parameters. Studying the properties of these intrinsic relaxations and linking them to their microscopic origins is essential to elucidate and understand the conduction mechanism in solids for their suitable multifunctional applications [1-6]. The measurement of different dynamical observables such as viscosity ($\eta$), dielectric relaxation time ($\tau$), conductivity ($\sigma$), magnetization (M) and polarization (P) as a function of thermodynamic intensive parameters (i.e., pressure (P), temperature (T)) deliver the quantitative and qualitative description of relaxation dynamics [1-5]. In this context, the Arrhenius law cannot describe satisfactorily the relaxation dynamics for many disordered materials. The interaction between the relaxing units/ species involved in thermally and/or pressure activated conductivity and dipolar process leads to the non-Arrhenius behavior described by Vogel-Fulcher-Tamann (VFT) relation and variable range hopping (VRH) model. Recently, ample progresses have been made to understand the effect of pressure (free volume) and temperature (thermal energy) on relaxation behavior of non-crystallizing liquids by different scientists around the globe. Nevertheless, the potential effects of pressure on relaxation dynamics in ferroelectric perovskites ($ABO_3$) have not been explored yet [1-6].

$BiFeO_3$ (BFO) is one of the most classical multiferroic research materials in condensed matter physics at present. It has high ferroelectric transition $T_C \approx 1103$ K and antiferromagnetic transition $T_N \approx 643$ K temperatures [7-12]. The origin of ferroelectricity in BFO is due to stereochemistry of Bi 6s orbitals while spin of transition metal cation $Fe^{3+}$ is responsible for G-type antiferromagnetic ordering with a long range spin cycloid modulation with period of 620–640 Å. In particular, a giant polarization $P_r$ at the level of 100 C/cm$^2$ has been detected in BFO thin films and single crystals, demonstrating that they are promising candidate materials for Lead-free ferroelectric applications [7-12]. Recent experimental investigations demonstrate that suitable chemical substitution/ doping efficiently tailor and enhance the multiferroic/ physical properties of BFO through chemical pressure (i.e., perturbation of local charge ordering) [13].

This is due to the possibility of driving the crystal symmetry of BFO towards a morphotropic phase boundary (MPB) and eventually improving its properties [13]. Aside from the importance of BFO in novel applications, a lesser attention has been devoted to the experimental investigations on the fundamental aspects of spin glass behavior, nearly constant loss (NCL) features, conductivity scaling and effect of high pressure[7, 14-22]. There are also still many open questions on the low temperature anomalies with magnetic origin. Perhaps the most widely studied phase transitions are those at T = 140 K and T = 200 K, examined independently by Cazayous *et al.* [14] and Singh *et al* [15-16]. Recently, Jarrier *et al.* reported that the T = 200 K anomaly in BFO is attributed to the onset of spin glass feature [23]. Redfern *et al.* described the anomalies in BFO at T = 200 K as magneto-elastic with small coupling to polarization [19].The coupling between magnetic, electric and elastic order parameters varies according to the length scale of the transition dynamics in each case: dipolar and strain coupling leads to the possibility of long-range correlations dominating the spin-glass behaviour, resulting in a rare example of an acentric mean-field spin glass in BFO [19]. The existence of intrinsic coupling between electric (dipole) and magnetic order (spin) in BFO offers a fundamental bridge to understand and resolute the magnetic spin glass behavior through the aspects of relaxation dynamics using dielectric relaxation spectroscopy. This experimental technique has been adopted widely to probe and understand the molecular dynamics of the glass forming systems in a large time window [1-6].

Much advance in understanding the ferroelectrics / multiferroics have also been achieved by investigating the role of temperature, electric field (E), magnetic field (H), synthesis technique and chemical composition (x) in BFO ceramics (bulk/ nano), single crystals and thin films [24-25]. In perovskite $ABO_3$-type materials, a potential connection exists between electrical properties and structural modification. Eventually, the dielectric relaxation depends on external thermodynamic variables due to electron-lattice coupling [26-27]. In this prospective, the external high-pressure can be considered as a suitable variable compared to other parameters since it acts only on interatomic distances [28-33]. By applying pressure to a sample of fixed composition, one varies only the interatomic interactions and balance between long- and short-range forces, making it simpler to uncover to the fundamental physics. In particular, the energetic order between different phases in ferroic materials can be notably perturbed by compression. This new and unexpected high pressure-ferroelectricity is different in nature from conventional ferroelectricity because it is driven by an original electronic effect rather by ionically-driven long-range interactions [28-33]. In the year 1975 Samara and co-workers studied first time role of pressure on $Pb_{(1-3x/2}La_xZr_{1-y}Ti_yO_3$ relaxor feroelectrics [29]. Later Molak and co-workers studied electrical properties of the $PbMn_{1/3}Nb_{2/3}O_3$ ceramics under pressure at isothermal and isobaric conditions using dielectric spectroscopy [34]. Aside from the above experimental work, any substantial investigation has not been carried out in BFO till date except the effect of pressure on crystal symmetry evolution by Kreisel and co-workers [28, 30-31]. In conclusion, it is essential to understand and establish the potential connection between chemical pressure induced by substitution (x) and external hydrostatic pressure (P).

In the quest to have a suitable material for examination, we have modified $BiFeO_3$ with 10% $GdMnO_3$ substitution with the aim to achieve a stabilized structure and enhanced magnetic properties that include mixture of ferro- and anti-ferromagnetic features (supplementary material) [7, 18, 58]. This elevated degree of magnetic frustration and the relative competition between ferro and anti-ferromagnetic interactions makes the modified material $Bi_{0.9}Gd_{0.1}Fe_{0.9}Mn_{0.1}O_3$ appropriate choice to highlight the glassy nature and associated features. The plan of the paper is as follows: (a) to understand the essential aspects of the relaxation dynamics and conduction mechanism in $ABO_3$-type multiferroic nanoceramic $Bi_{0.9}Gd_{0.1}Fe_{0.9}Mn_{0.1}O_3$ under wide range of temperature (153 K $\leq$ T $\leq$ 423 K; $\Delta$T = 5 K), external hydrostatic pressure (0.1 MPa $\leq$ P $\leq$ 1.7 GPa) and aging time (0 s $\leq t \leq$ 80000 s; $\Delta$t = 1000 s ), (b) to explore the glassy (non-equilibrium) state below T $\approx$ 200 K though aging experiment and the influence of nearly constant loss (NCL) feature on relaxation, (c) effect of time-temperature superposition using modulus scaling and (d) effect external high pressure and its connection to chemical pressure on the material. In this work, we aim to highlight the above mentioned fundamental issues using broadband dielectric spectroscopy and calorimetric study.

**Experimental details**

The details of sample synthesis and optimization have been reported in our previous works [35]. Dielectric spectroscopy: The dielectric spectra were collected over a wide frequency ($f = 10^{-2}$–$10^6$ Hz) and temperature range (T = 153 K to T = 423 K with step $\Delta$T = 5K) using a Novo-Control GMBH Alpha dielectric spectrometer. The temperature was controlled by the Novo-Control Quattro system, with the use of a nitrogen gas cryostat. Temperature stability was better than 0.1 K. Differential scanning calorimetry (DSC): Calorimetric measurements of examined DES were performed with a Mettler-Toledo DSC apparatus equipped with a liquid nitrogen cooling accessory and an HSS8 ceramic sensor (heat flux sensor with 120 thermocouples). Temperature and enthalpy calibrations were carried out by using indium and zinc standards. High pressure dielectric spectroscopy: For high pressure studies the sample capacitor was placed into a high-pressure stainless steel chamber and compressed using silicone fluid. Temperature was stabilized within 0.5 K via a Julabp thermostatic bath with circulating thermal fluid. The sample was investigated isothermally using a Novocontrol Alpha impedence analyzer up to P = 1.8 GPa. The pressure setup has been described elsewhere [36]. Aging experiment: We have carried out aging experiment above and below T = 200 K in two separate methods. The former method is connected with the measurement of dielectric data at individual isotherm below T = 200 K after recoiling from a reference isotherm T = 240 K (stable point/ equilibrium state) above transition. First the dielectric response was recorded at T = 240 K. Then the temperature was lowered to T = 200 K for the measurement. Again the system was heated to T= 240 and cooled to T = 190 K. The data was collected below T = 200 K down to 150 K with a step of 10 K with respect to a stable point at T = 240 K. The latter method is connected to direct time dependent measurement of relaxation time from $t$ = 0 s to $t$ = 80000 s in the frequency range ($f$ = $10^{-2}$ Hz to $10^{6}$ Hz). The data was collected for time interval Δt = 1000 s. X-ray diffraction: The X-ray diffraction (XRD) data were collected using X-ray powder diffractometer (PANalytical, PW3040, Netherlands) at very slow scan of 2° /min with a step size of 0.02° in a wide range of Bragg's angle 2θ (20° ≤2θ ≤ 80°) using Cu-K$\alpha_1$ radiation (λ = 1.5405 Å) with Nickel filter. FESEM Studies: The Nova Nano SEM 450 was used to study the surface morphology (grain size, void and uniformity of sample matrix) of the system. Magnetic study: The magnetic properties were determined by wide-range Superconducting Quantum Interference Device (SQUID) magnetometer MPMS XL7 Quantum Design, at temperature range of 2–400 K and magnetic field up to $\mu_0$H = 7 Tesla.

## Results and discussion

*Relaxation dynamics at ambient pressure*

In permittivity formalism, dielectric response $\varepsilon^* = \varepsilon'(f) + \varepsilon''(f)$ of the system indicate the oscillation of electric displacement vector (**D**) at constant electric field (**E**) where the real, $\varepsilon'(f)$ and imaginary, $\varepsilon''(f)$ components are the storage and loss of the energy during each cycle of the electric field, respectively. It represents the dielectric retardation process [37]. Figures .1 (a) and (b) illustrate frequency dependent $\varepsilon'(f)$ and $\varepsilon''(f)$ of $Bi_{0.9}Gd_{0.1}Fe_{0.9}Mn_{0.1}O_3$ for several isotherm between T = 423 K to T = 173 K at a step ΔT= 50 K at ambient pressure (P = $10^5$ Pa). It is observed that the magnitude of $\varepsilon'(f)$ decreases with increasing frequency at all temperatures which is a characteristic of polar dielectric material. A closer examination of $\varepsilon'(f)$ spectra reveals step-like frequency dependence and the departure of $\varepsilon''(f)$ from the $\upsilon^{-1}$ dependence. It clearly indicates the existence of relaxation process in the experimental window. The upward trend in ε″ (*f*) at low frequencies $f \approx 10^2$ Hz is credited entirely to dc-conductivity and electrode polarization. However, in the case of a semiconductor, analysis of relaxation processes in the permittivity representation is often difficult due to the significant dc-conductivity [1]. The large values of dielectric loss at low frequencies due to conduction mask any loss peaks. Consequently, we have adopted the derivative formalism to resolve relaxation processes in this temperature range using following relation; $\varepsilon''_{der} = -\frac{\pi}{2}\left[\frac{d\varepsilon'(\upsilon)}{d\log\upsilon}\right]$. Since the permittivity parts $\varepsilon'(f)$ and $\varepsilon''(f)$ are interrelated by the Kramers-Kronig transforms and both the quantities are equivalent with respect to their information [1]. As a result, we obtain conductivity-free dielectric loss peaks as depicted in figure. 1(c) and relaxation peak moves towards low frequencies with decreasing temperature. The thermal evolution of relaxation time, determined as the inverse of the maximum peak frequency, $\tau_{max} = \frac{1}{2\pi\upsilon}$ is displayed in the Arrhenius representation in figure. 3(d) by the equation $\log\tau = \log\tau_\infty + \frac{E_a}{RT}\log\tau_e$. The experimental data points fit to the Arrhenius law with $E_a$ = 0.36 eV (±0.009) and the characteristic time $\tau_0 \approx 10^{-9}$ s (figure.1 (d)).

It is well established that the electrical properties of perovskite type materials are closely related to its crystal structure and oxygen vacancies [26, 38]. The oxygen vacancies and other charge carriers (i.e., electrons and holes) can be generated gradually at various stages of calcination and sintering at high temperatures during the sample preparation. This process can be described by the Kröger-Vink notations as follows; $O_0 \rightarrow \frac{1}{2}O_2 + V_O^{\cdot\cdot} + 2e'$ where the two liberated electrons can be captured by the $Fe^{3+}$ and $Mn^{3+}$ ions. Likewise, charge transport by electron hopping takes place by means of an oxidation-reduction serial process between the Fe and Mn ions via the following structural chains of $Fe^{2+} - O - Fe^{3+}$and $Mn^{3+} - O - Mn^{4+}$ or combinations. More importantly electron hopping, when resulting from the evacuation of a lattice, leaves behind a charge imbalance, creating a virtual dipole. This dipole then creates a form of dipole-lattice interaction, which induces relaxations [27]. In addition, the contribution of the conduction (electrons or holes or protons) to dielectric polarization has been reported in many systems for single crystals and polycrystalline ceramics [26, 39-41]. The polarization was greatly enhanced by the interaction of the electrons (created by the ionization of oxygen vacancies) and the dielectric relaxation process. In this case, even with low concentration of the dipoles, the system could display a very magnitude of high permittivity. In the present examined material, we detect a high value of $\varepsilon'(\upsilon)$ below $f$ = $10^4$ Hz. The magnitude of activation energy $E_a$ = 0.36 eV (±0.009) from thermal dependence of relaxation time correspond to the combination effect of the reorientation of the off-center $A^{3+}$ and $B^{3+}$ ions coupling in $ABO_3$ with the conducting electrons *i.e.,* electron and/or hole-phonon interaction. This behavior has been explained as the evidence of small polaron hopping (electron-lattice coupling) for several single crystals [26, 42].

The relaxation times estimated from loss spectra have a macroscopic meaning because they are associated with macroscopic polarization decay of sample and the macroscopic relaxation times ($\tau_{max}$) connect to the microscopic times ($\tau_{micro}$) of Debye model through the relation; $\tau_{max} = \tau_{micro}\left[\frac{(\varepsilon_0+2)}{(\varepsilon_\infty+2)}\right]$. Hence, $\tau_{micro}$ in $ABO_3$-type oxides are likely to display $10^1$–$10^2$ times shorter than those determined directly from dielectric loss peaks because of high magnitude of permittivity [1, 34]. For systems with significant electric conduction that covers or strongly overlaps with the other relaxation process, the dielectric response can be also examined by electric modulus representation, which suppresses the dc contribution $M^*(f) = \frac{1}{\varepsilon^*(f)} = \frac{i2\pi\upsilon\varepsilon_0}{\sigma^*(f)}$. This approach is useful for various conducting materials such as ionic conductors and polycrystalline materials. Physically, the electric modulus describes the relaxation of electric field (**E**) in the material sample when the electric displacement (**D**) remains constant. Ultimately, the modulus representation describes the real dielectric relaxation process. For Debye process, the relaxation times $\tau_M$ are related to the time determined from dielectric permittivity $\tau_\varepsilon$ by the relation $\tau_\varepsilon = \tau_M\left(\frac{\varepsilon_0}{\varepsilon_\infty}\right)$. Hence, the values of relaxation times obtained from electric modulus coincide with the relaxation times in the microscopic meaning [1, 34].

Figure.2 represents the frequency dependent ($f$ = $10^{-1}$ to $10^{6}$ Hz) imaginary part of dielectric modulus $M''(f)$ at ambient pressure (P = $10^{5}$ Pa) in the temperature range (163 K ≤ T ≤ 423 K: Δ T = 20 K). Three relaxation peaks are clearly evident in the experimental window. For $M''(\upsilon)$ at T = 423 K, it is clear that first relaxation (denoted as $\tau_{\alpha}$) appears near $f_{max} \approx 10^{6}$ Hz and the second (denoted as $\tau_{\sigma}$) one close to $f_{max} \approx 10^{4}$ Hz. On decreasing temperature both the relaxation moves with different thermal activation towards low frequency side. The appearance and enhancement of new relaxation (denoted as $\tau_{\beta}$) takes place at high frequency below T = 200 K. The asymmetric power-law behavior of the modulus data at low and high frequencies as shown in figure. 2 suggests that relaxation process can be properly examined by the Havriliak-Negami (HN) relaxation function, $F^{*}_{HN}(f)$ which takes the mathematical form $F^{*}_{HN}(f) = \frac{1}{[1+(if\tau_{HN})^{\alpha}]^{\beta}}$, where $\tau_{HN}$ defines a characteristic relaxation time and the shape parameters, α and β are related to (1-$n$) = $\beta_{KWW}$ by the equation, $\alpha\beta = (\beta_{KWW})^{1.23}$ [1, 43]. Using sum of three HN fittings, we have calculated the following dynamic parameters from the $M''(f,T)$ spectra: (a) the relaxation times ($\tau_{\alpha}$, $\tau_{\sigma}$ *and* $\tau_{\beta}$) corresponding to $M''(f)$ maxima; $\frac{1}{2\pi f_{max}}$ , (b) the shape parameters. Using the above information, we have constructed the relaxation map of the examined material by plotting $\log_{10}\tau$ function of $10^{3}$/T. This relaxation map mirrors the dynamic features of the system and central feature of our investigation (figure.3).

The relaxation map clearly indicates that the thermal activation of $\tau_{\alpha}$ relaxation spans through seven decades in frequency by non-expontial relaxation process described by Kohlrausch-Williams-Watts function with exponent $\beta_{KWW}$ = 0.5. A non-exponential type of conductivity relaxation suggests that ion migration takes place via hopping accompanied by a consequential time dependent mobility of other charge carriers of the same type in the vicinity [44]. Again, the shape clearly indicates that the motion of the charges leads not only to conduction but also substantial polarization. It is a typical behavior of localized polarons [45]. The formation of small polarons because of the fact that these oxides contain mixed-valence transition metal ions (TMI) and hopping/ tunnelling from the higher to the lower valence states of the TMI occurs [46]. The $\tau_{\alpha}(T)$ dependence exhibited a clear convex curvature *i.e.,* neither the classical Arrhenius nor the VFT-type behavior [34]. Despite of the curvature, it can be modeled in the first approximation with the Arrhenius relation and the experimental data below and above T ≈ 200 K has been fitted accordingly. The activation energy corresponding to temperature range 423 K ≤ T $\leq$ 200 K is equal to $E_a$ = 0.38 eV (±0.002) with characteristic relaxation time $\tau_0 \approx 10^{-11}$ s whereas for 200 K ≤ T ≤ 153 K range $\tau_\alpha$ has $E_a$ = 0.29 eV (±0.003) with $\tau_0 \approx 10^{-9}$ s. The reciprocal of the time constant above T = 200 K compares well with the frequency of the lattice for ionic solids over the frequency range of $f \approx 10^{11}$ Hz. These parameters are indicative of ionic jump or dipole relaxation [47]. Eventually, conduction is attributed to the localized orientation hopping of electron back and forth between two charge defects and the presence of induced or permanent dipoles assisted by small polaron mechanism [44]. It is a typical feature of hopping conduction inside grain/ bulk. It has been reported that activation energy higher than $E_a$ = 1 eV is connected to double ionised vacancies, while those single ionised are involved in a process characterized by tenth of eV [34]. Taking into account that both dielectric relaxation and electric modulus formalism display similar magnitude of activation energy 0.36 eV and 0.38eV respectively within low temperature range of investigation, one deduces that single-ionised oxygen vacancies are involved in this process. Therefore, the mixed ionic–polaronic conductivity is expected in this case as it includes charge carriers generated from the vacancies. A short-range hopping of the single ionized oxygen ions, similar to dipole reorientation, via energy absorption and loss, leads to the observed anomaly in dielectric permittivity [34]. In order to understand the curvature in $\tau_\alpha(T)$ dependence and departure from Arrhenius relation above T = 200 K, we have performed calorimetric studies. It is clear that DSC traces confirms a typical glass like transition near T ≈ 200 K as observed for typical vitrious ionic conductors/ structural glasses and also anomaly in temperature dependent $\beta_{KWW}$ (inset figure.3). Recent investigations reports that T = 200 K is connected to the beginning of spin glass nature in BFO [23] and ultimately, the present experimental data indicates a strong intrinsic magneto-electric coupling. The analysis of convex curvature in $\tau_\alpha(T)$ will be discussed later.

The second relaxation $\tau_\sigma$ has thermal activation energy $E_a$ = 0.25 eV(±0.006) with $\tau_0 \approx 10^{-9}$ s in the temperature range 400 K ≤ T ≤ 285 K. It should be noted that this magnitude of activation energy is comparable to the value of $E_a$ = 0.29 eV for the relaxation in $LuFe_2O_4$ ceramic due to a two-site polaron hopping process of the charge transfer between $Fe^{2+}$ and $Fe^{3+}$ or Mn combinations [48-49]. Taking into account the giant low-frequency dielectric response and activation energy, the relaxation most likely arises from the inhomogeneous structure *i.e.,* the grain boundary.

The activation energy of fastest relaxation appearing below spin glass transition is equal to $E_a$ = 0.16 eV (±0.014) with $\tau_0 \approx 10^{-8}$ s. This value of activation energy well compares with the $E_a$ = 0.1 eV of the first-ionization of oxygen vacancies. Figure.4 compares the frequency dependent ac conductivity ($\log \sigma_{ac}$) for several temperatures in range 153 K ≤ T ≤ 173 K at step ΔT = 5 K. It is observed that for T = 153 K, ac conductivity varies linearly above a characteristic frequency $f$ = $10^4$ Hz *i.e.,* $\sigma = \sigma_0 + Af^n; n = 1$, which indicate the appearance of nearly constant loss feature (NCL). The linear dependence in $\log \sigma_{ac}$ as a function of frequency consents to estimate the crossover frequency (hopping to cage) that moves towards high frequency side on increasing temperature. The crossover points establish an Arrhenius type thermal activation with energy $E_a$ = 0.17 eV (±0.015) as shown in inset figure.4. More intriguingly, the same value is obtained from spin-lattice and electrical conductivity relaxation, for the barrier for short-range ion motion, $E_m$ = 0.17 eV *i.e.,* the height of the single ion potential well [50]. Again, it provides strong evidence that the fast relaxation below spin glass transition has potential connection with NCL feature (vibrational character caused by anharmonicity) [50] and originates from cage to short range hopping motion of electron in following structural chains of $Fe^{2+} - O - Fe^{3+}$and $Mn^{3+} - O - Mn^{4+}$ and combinations[27]. Moreover, it has to be emphasized that the behavior of this fastest relaxation and its connection to NCL feature in the present sample has never reported before.

Our experimental data clearly shows that the thermal activation of $\tau_\alpha$ display a deviation from Arrhenius type below T = 200 K with resolute convex curvature and indicate a temperature dependent activation energy. Eventually, it is expected that activation energy; $E_a = KT\left(\frac{T}{T_0}\right)^p$, governed by the dimensionality ($p$) of conduction. The exponent value is equal to $p$ =1/4 for the three dimensional case of conductivity process assuming a disordered crystal structure often observed in $ABO_3$-type materials [34]. Thus the variable range hopping (VRH) of small polaron model was checked to realize the possible conductivity mechanism that has been proposed for disordered solids at low temperatures. It is well known that several perovskite manganites exhibit electric conductivity described in terms of the VRH model of the small polaron; $\tau(T) = \tau_0 \exp\left(\frac{T_0}{T}\right)^{1/4}$ that can be related to the Fermi-glass features as observed in magnetic experiments, where the parameter $T_0$ is related to the disorder energy which corresponds to variation in the local environment of the crystal lattice sites participating in the hopping process [34]. Most strikingly, it is found that the experimental data consents with VRH model satisfactorily with $T_0 = 4.6 \times 10^9$ K. In order to comprehend and provide a direct evidence of non-equilibrium state below T = 200 K, we have carried out aging experiment for $\tau_\alpha$ above and below T = 200 K in two separate methods (figure. 3 and figure.5). The former method is associated with the measurement of dielectric responses at individual isotherm below T = 200 K after recoiling from a reference isotherm each time (stable point/ equilibrium state T = 240 K) above transition whereas the latter method is connected to direct time dependent measurement of $\tau_\alpha\,(t)$ from $t$ = 0 s to $t$ = 80000 s in the frequency range ($f$ = $10^{-2}$ Hz to $10^6$ Hz). The details of the experiment have been described in experimental section. II. In the former one, it is established that relaxation time is unaffected upon aging at T = 240 K whereas relaxation time decreases slightly below transition and activation energy drops only by $\Delta E_a$ = 0.02 eV compared to the unaged $\tau_\alpha\,(T = 240\ K, t = 0\ s)$. It indicates that $\tau_\alpha$ is not actively sensitive to the former approach. In the latter approach, we notice that the peak position and $\beta_{KWW}$ is not sensitive to aging (even up to $t$ = 80000 s) at isotherm T = 240 K and eventually provides the evidence of equilibrium state as observed above. But for the isotherms at T = 165 K and T = 178 K, we perceive an ample decrease in $\tau_\alpha$ relaxation time (or increase in conductivity) of magnitude $\Delta f$ = 0.1 and $\Delta f$ = 0.2 decade respectively as shown in figure.5. Again, it should be noted that peak height $M''_{max}$ increases and width gradually decreases upon aging [51]. The present experimental data provides the strong evidence for the non-equilibrium state (glassy nature) of the examined sample below T = 200 K through dielectric spectroscopy for the first time. Again, it strongly supports conduction mechanism through VRH model and the glassy nature of the sample.

In order to monitor the shape of the relaxation profile as a function of temperature, we have constructed the master curve by superimposing a number of dielectric modulus profiles recorded at different temperatures (i.e., above and below spin glass transition T=200 K), to the spectrum obtained at T = 243 K as the reference as shown in figure.6. The obtained master plot clearly shows that the shape of the conductivity-profile is practically invariant upon heating. This observation provides the evidence to support the time-temperature superposition in the present examined materials. Again, it implies that conduction mechanism is independent of temperature.

Moreover, the dynamic processes of the relaxing species/ units occurring at different time scales, the distribution of the relaxation times is temperature independent [2, 38,43].

In the previous discussions, the dielectric relaxation behaviors have been examined in the formalisms of modulus ($M^*$) and permittivity ($\varepsilon^*$) where localized movement of carriers is dominant. In the case of long range movement, the resistive and conductive behaviors are often analyzed by impedance ($Z^*$). Sometimes it is difficult to understand the data whether the electrical response is due to long-range conductivity (de-localized) or dipole relaxation (localized) of materials. Both localized and delocalized conductions are bulk processes, and therefore this gives rise to the similar magnitude of geometrical capacitance [27, 38, 42, 52-56]. The method of impedance spectroscopy allows us to distinguish between grain and boundary effect because they have different magnitude of relaxation times in frequency spectrum. The $Z''(f)$ spectra signifies resistive effect whereas $M''(f)$ amplifies capacitive effect. Consequently, the combined plot of $M''(f)$ and $Z''(f)$ vs. frequency is able to distinguish whether the short-range or long-range movement of charge carries is dominant in a relaxation process [27, 38, 42, 52-56]. The separation of peak frequencies between $M''(f)$ and $Z''(f)$ indicates that the relaxation process is dominated by the short-range movement of charge carriers and departs from an ideal Debye-type behavior while the frequencies coincidence suggests that long rang movement of charge carriers is dominant. Figure. 7(a) shows the combined plot of frequency dependent $M''(f)$ and $Z''(f)$ spectra at T = 348 K. As can be seen, the M″( *f*) and Z″( *f*) peaks are not coinciding indicating the departure from Debye relaxation in the sample. The departure from the ideality justifies the presence of constant phase element (CPE) in the equivalent circuit proposed for the sample. Impedance of CPE is given by $Z_{CPE} = [A_0(i\omega)^n]^{-1}$, where $A_0 = \frac{A}{\cos\left(\frac{n\pi}{2}\right)}$, A and *n* are frequency independent but depends on temperature [27, 38, 42, 52-56]. The *n* value lies between 0 and 1 (*n*=1 for an ideal capacitor and *n*=0 for an ideal resistor).

The complex impedance spectra $Z'$vs. $Z''$ (Nyquist-plots) of the examined sample measured at different temperatures are shown in figure. 7(b). The complex impedance plot typically comprises of two overlapping semicircular arc at low temperatures with center below the real axis suggesting the departure from ideal Debye behavior. At low temperatures (below T = 163 K not shown), a straight line is observed in impedance plot signifying the insulating property of the material. Above T = 200 K, a semicircular arc starts forming and finally semicircular arcs become prominent on increasing temperature. The presence of a single semicircular arc is due to the bulk property of the material. At higher temperature (T = 263 K) the plot can be characterized by the presence of two clear overlapping semicircular arcs with their centers below the real axis. The high frequency semicircle is attributed to the bulk (grain) property of the material, whereas the low frequency arcs (at high temperature) are ascribed to the presence of grain boundary effect. The intercept of the semicircular arcs on the real axis gives rise to bulk and grain boundary resistance of the materials. The bulk and grain boundary resistance (d.c.) of the samples decreases with rise in temperature as indicated by the corresponding drop in the diameter of the semicircles on increasing temperature. The assignment of the two semicircular arcs to the electrical response due to grain interior and grain boundary is consistent with the brick-layer model for a polycrystalline material [27, 38, 42, 52-56]. So the high frequency semicircular arc can be modeled to an equivalent circuit of parallel combination of a resistance (bulk resistance), capacitance (bulk capacitance) along with a constant phase element, whereas the low frequency semicircular arc can be modeled for parallel combination of a resistance (grain boundary resistance) and a capacitance (grain boundary capacitance). Both the equivalent circuit corresponding to grain and grain boundary contribution are connected in series for fitting the impedance data as shown in figure.7 (a). The impedance data (symbols) have been fitted (solid line) with the above proposed model by the commercially available software ZSIMP-WIN as shown in figure. 7(b).

The d.c. conductivity due to the bulk and grain boundary contributions have calculated from the resistances obtained after the fitting the impedance data using the formula, $\sigma_{dc=}\frac{l}{RA}$, where $l$ = thickness, A = area and R = resistance. Figure.7(c) shows the variation of dc conductivity ($\log \sigma_{dc}$) (due to bulk and grain boundary) with inverse of absolute temperature ($10^3$/T). It has been observed that the d. c. conductivity decreases with rise in temperature suggesting the negative-temperature coefficient of resistance behavior, typically observed in semiconductors. The temperature dependence of bulk and grain boundary conductivity can be explained by the Arrhenius relation with activation energy $E_a$ = 0.39 eV (±0.005) and $E_a$ = 0.28 eV (±0.009) respectively below T = 200 K. Above that a change of slope takes place in bulk conductivity and drop in activation energy $\Delta E_a$= 0.09 eV takes place. This change in slope is attributed to the onset of spin glass transition. It is clear that the value of activation energy for both grain and grain boundary is nearly similar magnitude. It can be conclude that conducting species / relaxing species for both grain and grain boundary are the same or have similar origin.

*Relaxation dynamics under high pressure*

Figure.8 (a) displays the $log_{10}\ \varepsilon''(der)$ as function of frequency ($f$ = $10^{-1}$ to $10^{6}$ Hz) for several hydrostatic pressures ranging from P = 163 MPa to P = 1765 MPa at isotherm T = 263 K above spin glass transition. We notice that both the relaxation time $\tau_\alpha$ and peak width decrease upon compression. It indicates that the applied hydrostatic pressure reduced the heights of the energy barriers. On the contrary, such a feature in relaxation behavior observed in dielectric response is completely different from usually observed for glass-forming supercooled liquids, for which compression induces the slowing of the relaxation dynamics being related to the increase in the activation energy [34, 6]. Again, it is very important to note that high pressure significantly suppresses the electrode polarization which is evident from the decrease in dc conductivity at low frequencies. Intriguingly, the dielectric data at high pressure is a signature of intrinsic bulk property of the material.

Figure. 8(b) explains the isothermal pressure activation of $log_{10}\ \tau_\alpha$ . It is observed that relaxation time drops linearly upon increasing pressure up to P = 600 MPa and simple volume activated law [1,6] : $\tau = \tau_0 exp^{\left(\frac{P\Delta V}{RT}\right)}$ describe satisfactorily the experimental data points. The estimated activation volume in this pressure range P = 163 MPa to P = 496 MPa is equal to $\Delta V$ = -2.763 $cm^3$/mol. Above P = 600 MPa (approx.), relaxation times $\tau_\alpha$ decouples from linear pressure evolution and enters into non-linear pressure dependency (in the range P = 894 MPa to P = 1765 MPa) governed appropriately by VFT pressure law [1,6]: $\tau = \tau_0 exp^{\left(\frac{CP}{P_0 - P}\right)}$. This region has activation volume of magnitude $\Delta V$ = -1.761 $cm^3$/mol (liner approx.). The activation volume is a very useful parameter to characterize the relaxation processes in non-crystallizing liquids. Its value provides information on the pressure sensitivity to the relaxation times [34,6]. Thus, the value of $\Delta V$ may reflect on the volume requirements for local molecular motion of relaxing units. The natural consequence is that the size of a relaxing molecular unit will have an influence on the value of $\Delta V$. The decoupling from Arrhenius to VFT nature marks a cross over pressure (600 MPa) for the examined sample as shown in inset figure. 8(b). The pressure dependence of activation volume determined from the relation: $\Delta V = RT\left(\frac{d\,ln\,\tau_\alpha}{dP}\right)_T$ is shown in inset figure. 8(b). Physically, it implies that the time evolution of relaxing units/ species is cooperative above this critical pressure and pressure VFT relation takes care of the experimental data points satisfactorily. Again, the gradual narrowing of the $\tau_\alpha$ relaxation and decrease in activation volume above critical pressure strongly support the increase in intermolecular interaction which is translated as a decoupling feature with non-linear pressure dependence. As reported by Molak *et al.*, this cross over pressure indicates the structural transformation in the material [34]. On the contrary, Haumont *et al.*, observed a crystallographic change in $BiFeO_3$ only after P = 3 GPa. Thus, there exists no connection of crossover pressure to structural transformation [33]. Literature reports that suitable chemical substitution induces chemical pressure inside the material which leads to structural transformation. Gavriliuk *et al.* reported that the pseudocubic lattice parameter decreases at a rate of approximately −0.01 Å/GPa [57]. Structural investigation indicate that pseudocubic lattice parameter changes $\Delta a_p$ = 0.03 Å upon 10% substitution in parent $BiFeO_3$ (see supplemtary material) [35, 58]. This observation in chemical pressure strongly supports the high pressure investigations of structural transformation by Haumont *et al* [33]. It implies that intriguingly, there exists a parallel connection between chemical pressure driven by chemical substitution and external hydrostatic pressure.

In ferroelectrics, relaxor behavior origins from either frustration or compositionally induced disorder. Samara *et al.* observed a cross over from normal ferroelectric to relaxor by application of pressure below P = 1500 MPa in $Pb_{(1-3x/2}La_xZr_{1-y}Ti_yO_3$. They explained this behavior a large decrease in the correlation radius among polar nanodomains which is a unique property of soft ferroelectric phonon mode systems [29]. More intriguingly, the present experimental observation provides a strong evidence of pressure induced relaxor behavior of the examined sample through relaxation dynamics and have fundamental importance in ferroelectric relaxor physics. The relative competition between long range and short ranges forces in the material deliver the relaxor aspect through high pressure. This novel high pressure effects is different in nature compared to conventional ferroelectricity because it is driven by an electronic effect rather by ionically-driven long-range interactions.

## Summary and conclusions

On the basis of dielectric data which provides the direct access to the dynamic features of relaxation and conduction, collected for the present multiferroic sample, we have drawn the following conclusions. Complex modulus spectroscopy is an appropriate representation of dielectric data to understand the key features of conductivity relaxation and provide the potential connection to understand the microscopic origin of the process. Variable range hopping model of small polarons have been verified for bulk conductivity relaxation. Direct evidence non-equilibrium state (glassy nature) below $T \approx 200$ K was witnessed and verified in the sample by aging experiments. Signature of magneto-electric coupling was demonstrated. A potential connection between nearly constant loss feature and fastest relaxation appearing below non-equilibrium state was established. Modulus scaling supports the time-temperature superposition in the material and indicates that conduction mechanism is independent of temperature. Impedance spectroscopy separates the grain and grain boundary contributions. It indicates that non-localized process dominates the relaxation and conducting species / relaxing species for both grain and grain boundary are the same or have similar origin. High pressure studies provide strong evidence in the support of the pressure induced relaxor transition in the material. There exists a direct connection between chemical pressure induced by chemical substitution and external hydrostatic pressure.

**Acknowledgments:** Authors are deeply grateful for the financial support by the National Science Centre within the framework of the Maestro2 project (Grant No. DEC-2012/04/A/ST3/00337).

## Figure Captions

Figure.1 (color online) Frequency dependent ($f$ = $10^{-1}$ to $10^{6}$ Hz) behavior of **(a)** $log_{10}\varepsilon'(\upsilon)$, **(b)** $log_{10}\varepsilon''(\upsilon)$ and **(c)** $log_{10}\ \varepsilon''(der)$ of $Bi_{0.9}Gd_{0.1}Fe_{0.9}Mn_{0.1}O_3$ for several isotherm between T = 423 K to T = 173 K at a step ΔT= 50 K at ambient pressure (P = $10^{5}$ Pa). Not all spectra are shown **(d)** Activation plot from permittivity representation.

Figure.2 (color online) Frequency dependent ($f$ = $10^{-1}$ to $10^{6}$ Hz) imaginary part of dielectric modulus $M''(\upsilon)$ at ambient pressure (P = $10^{5}$ Pa) in the temperature range (163 K ≤ T ≤ 423 K: Δ T = 20 K).

Figure.3 (color online) The relaxation map for the examined material. Inset-DSC thermogram with temperature dependent $\beta_{KWW}$ signifying glass transition.

Figure.4 (color online) Frequency dependent ac conductivity ($\log\sigma_{ac}$) for several temperatures below glass transition in range (153 K ≤ T ≤ 173 K). Inset- activation plot of frequency corresponding to onset of NCL.

Figure.5 (color online) Time dependent dielectric response (aging experiment) on $\tau_{\alpha}$ at isotherm T = 240 K, T= 178 K and T = 165 K.

Figure.6 (color online) Master curve: Superimposed dielectric modulus spectra of the sample, at different temperatures above and below spin glass transition with reference to T = 243 K. The solid red line represents the HN fit.

Figure.7 (color online) **(a)** Combined plot of frequency dependent $M''(\upsilon)$ and $Z''(\upsilon)$ spectra at T = 348 K. **(b)** The complex impedance spectra $Z'$vs. $Z''$ (Nyquist-plots) of the examined sample measured at different temperatures. **(c)** Thermal activation of grain and grain boundary conductivity.

Figure.8 (color online) **(a)** The variation of $log_{10}\ \varepsilon''(der)$ as function of frequency ($f$ = $10^{-1}$ to $10^{6}$ Hz) for several hydrostatic pressures ranging from P = 163 MPa to P = 1765 MPa at isotherm T = 263 K above spin glass transition. **(b)** Pressure activation of relaxation time. Inset-Pressure dependence of activation volume.

Figure Captions (Supplemtary materials)

Figure.S1 (color online) (a) Pseudo-cubic lattice parameter as function of composition (x) and (b) SEM micrograph for $Bi_{0.9}Gd_{0.1}Fe_{0.9}Mn_{0.1}O_3$ at room temperature.

Figure.S2 (color online) M vs. H curves for $BiFeO_3$ and $Bi_{0.9}Gd_{0.1}Fe_{0.9}Mn_{0.1}O_3$ at T = 300 K and T = 2 K.

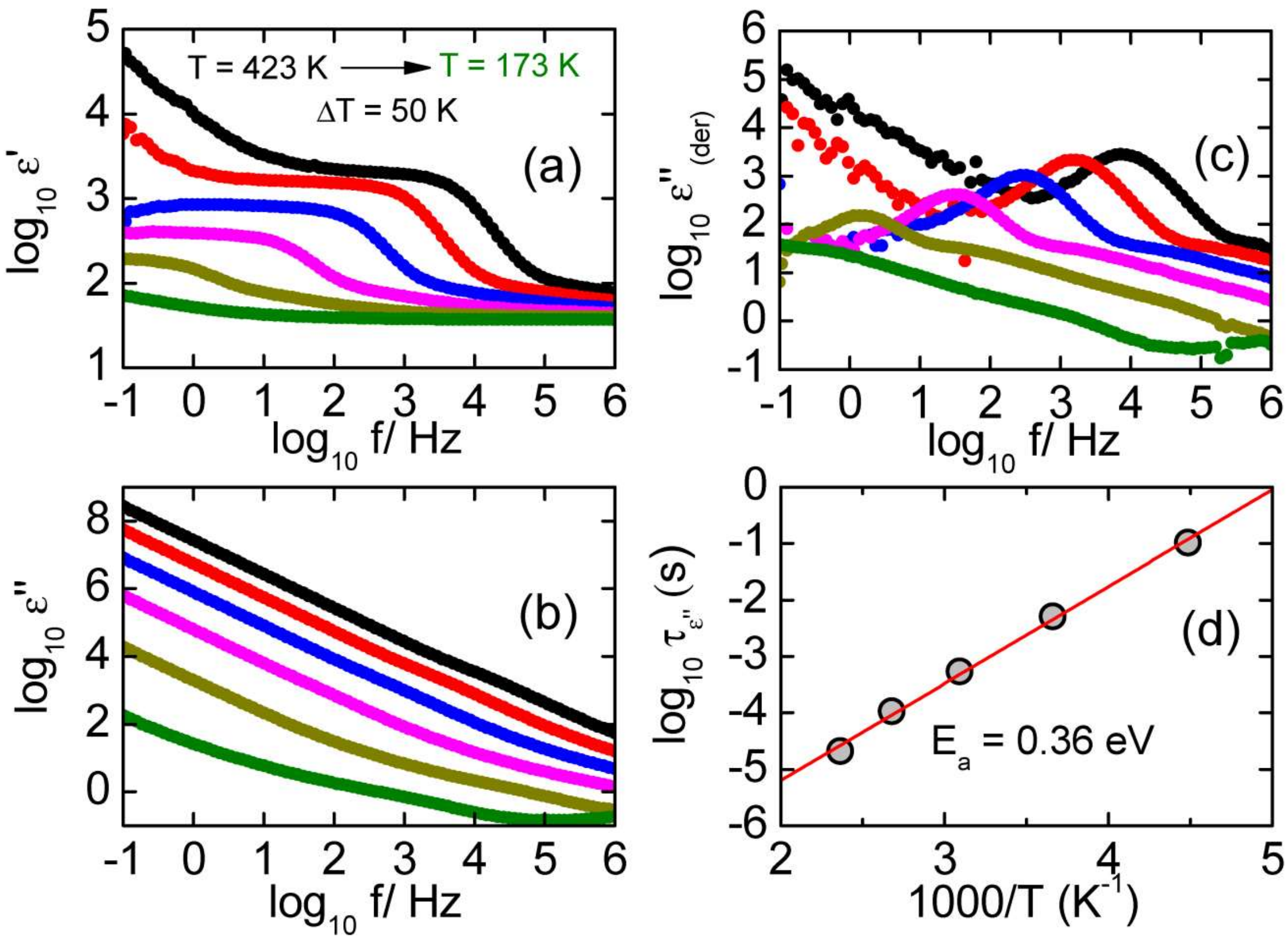

**Figure.1**

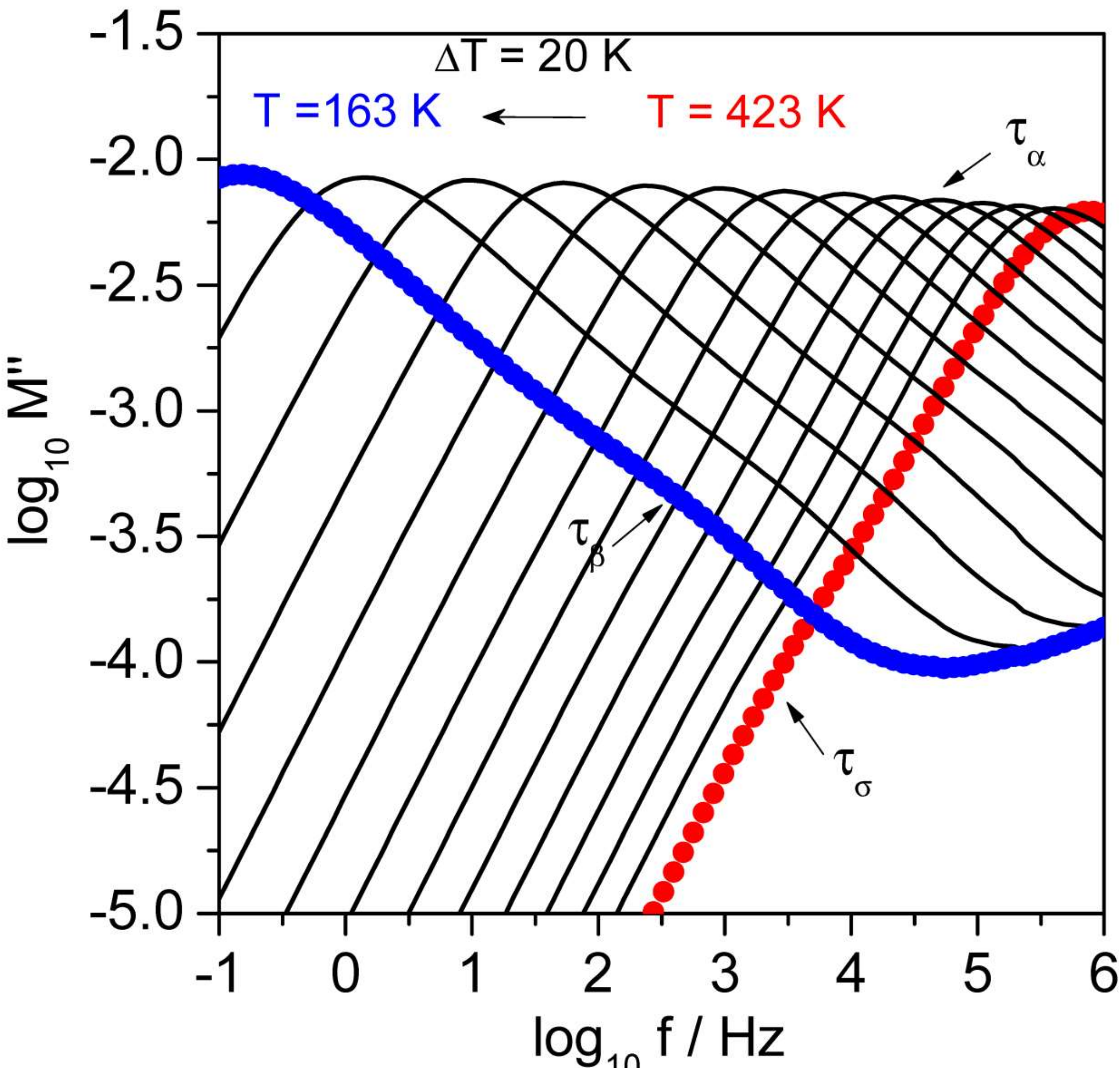

**Figure.2**

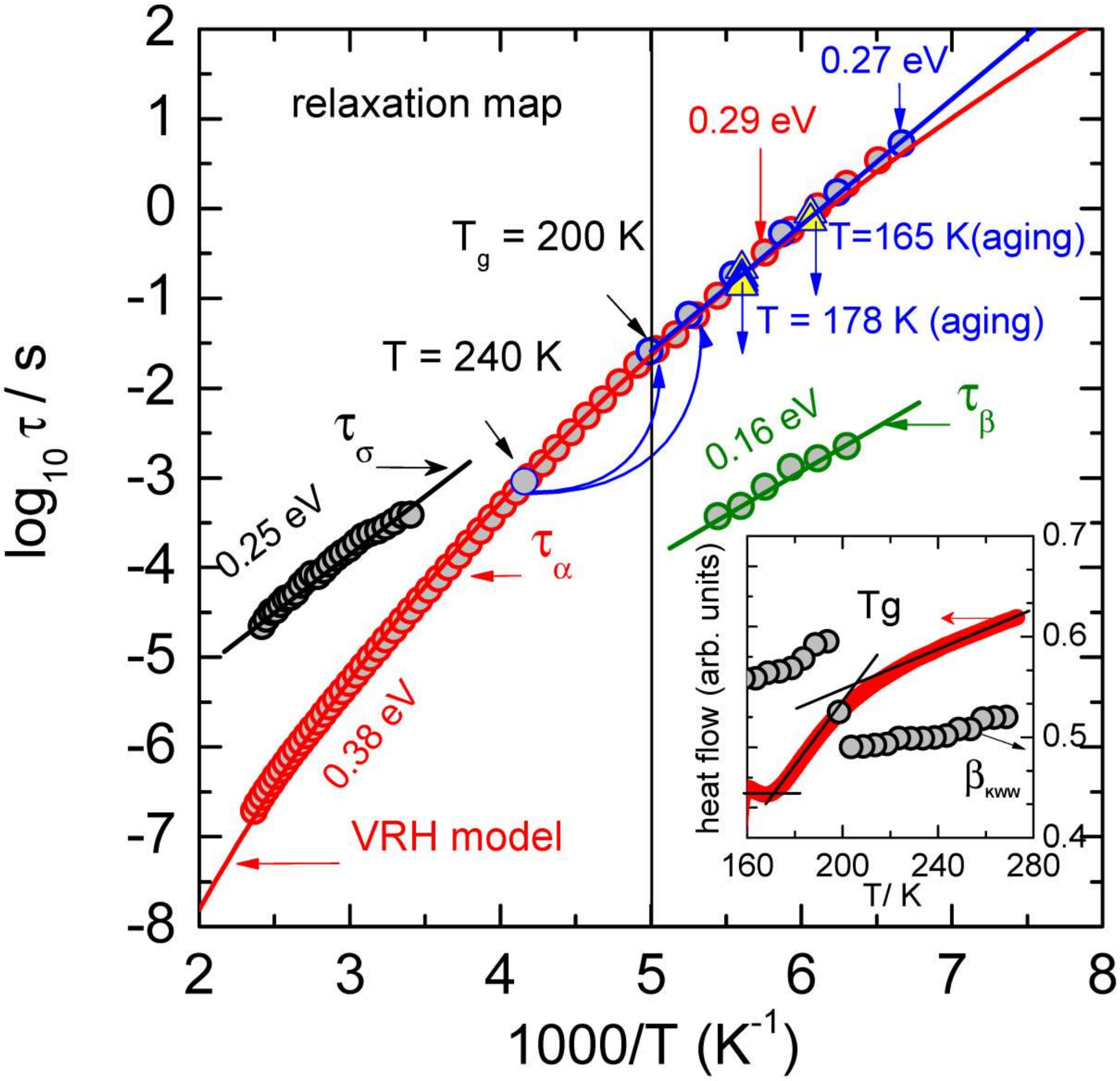

**Figure.3**

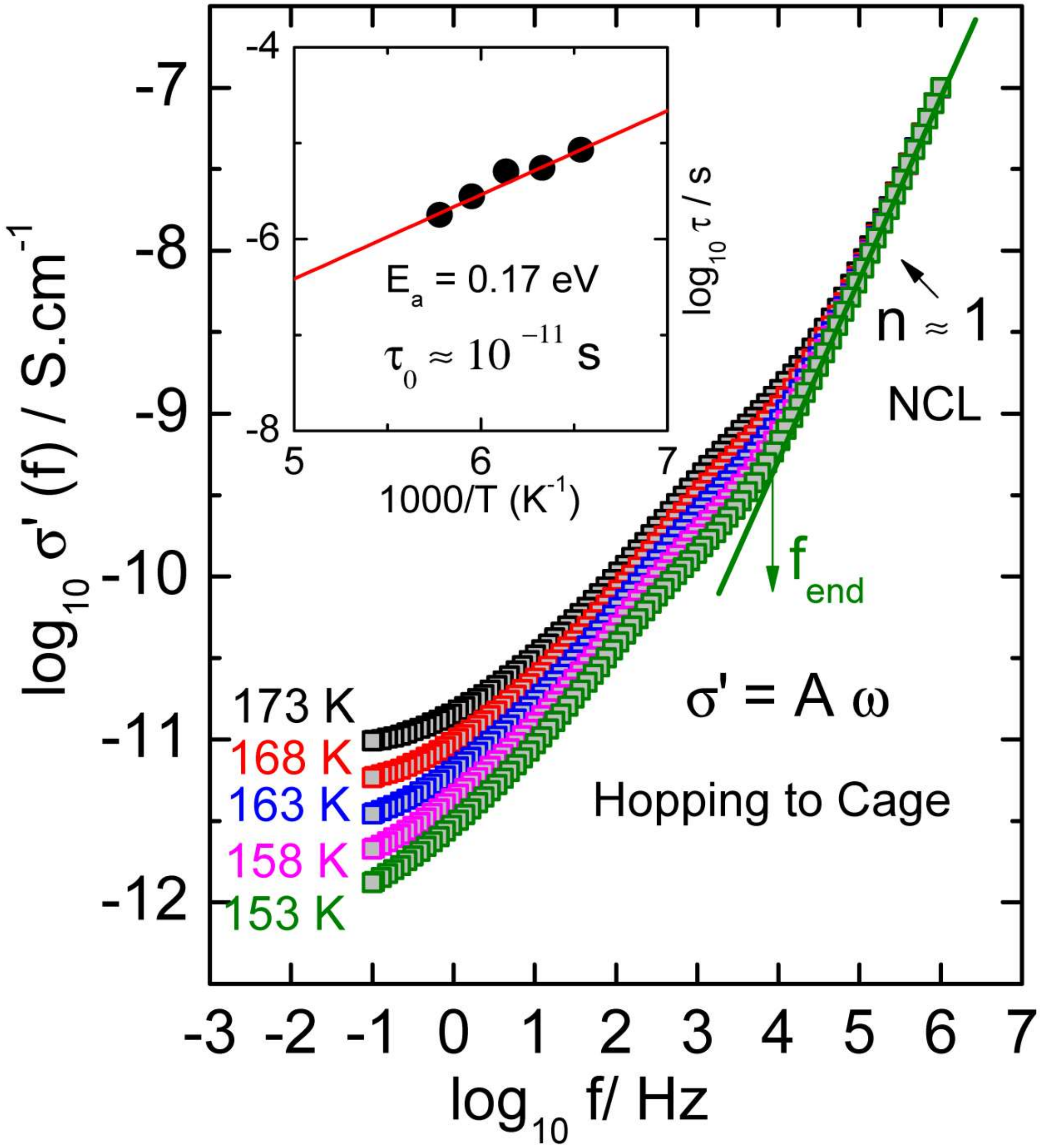

**Figure.4**

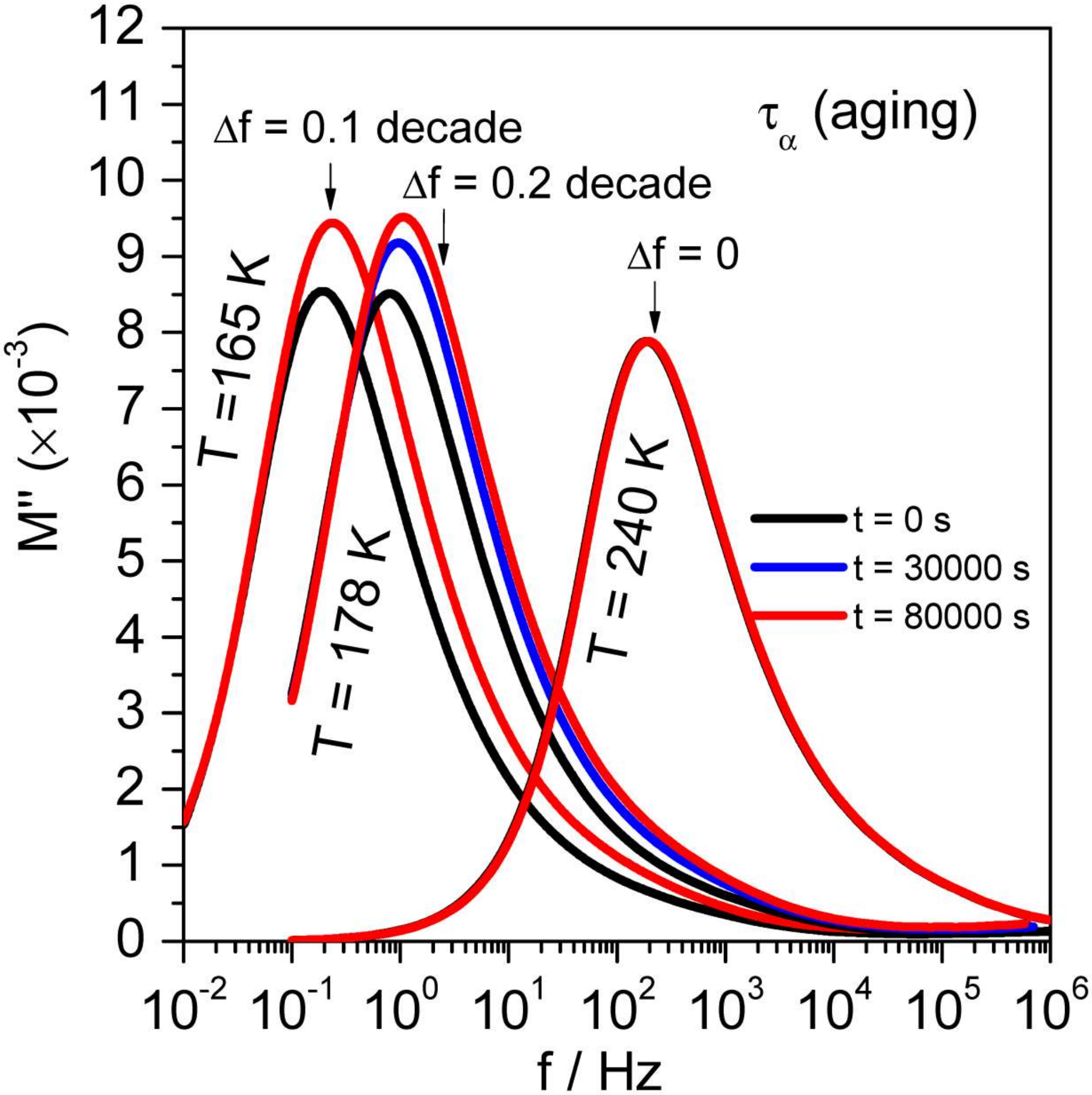

**Figure.5**

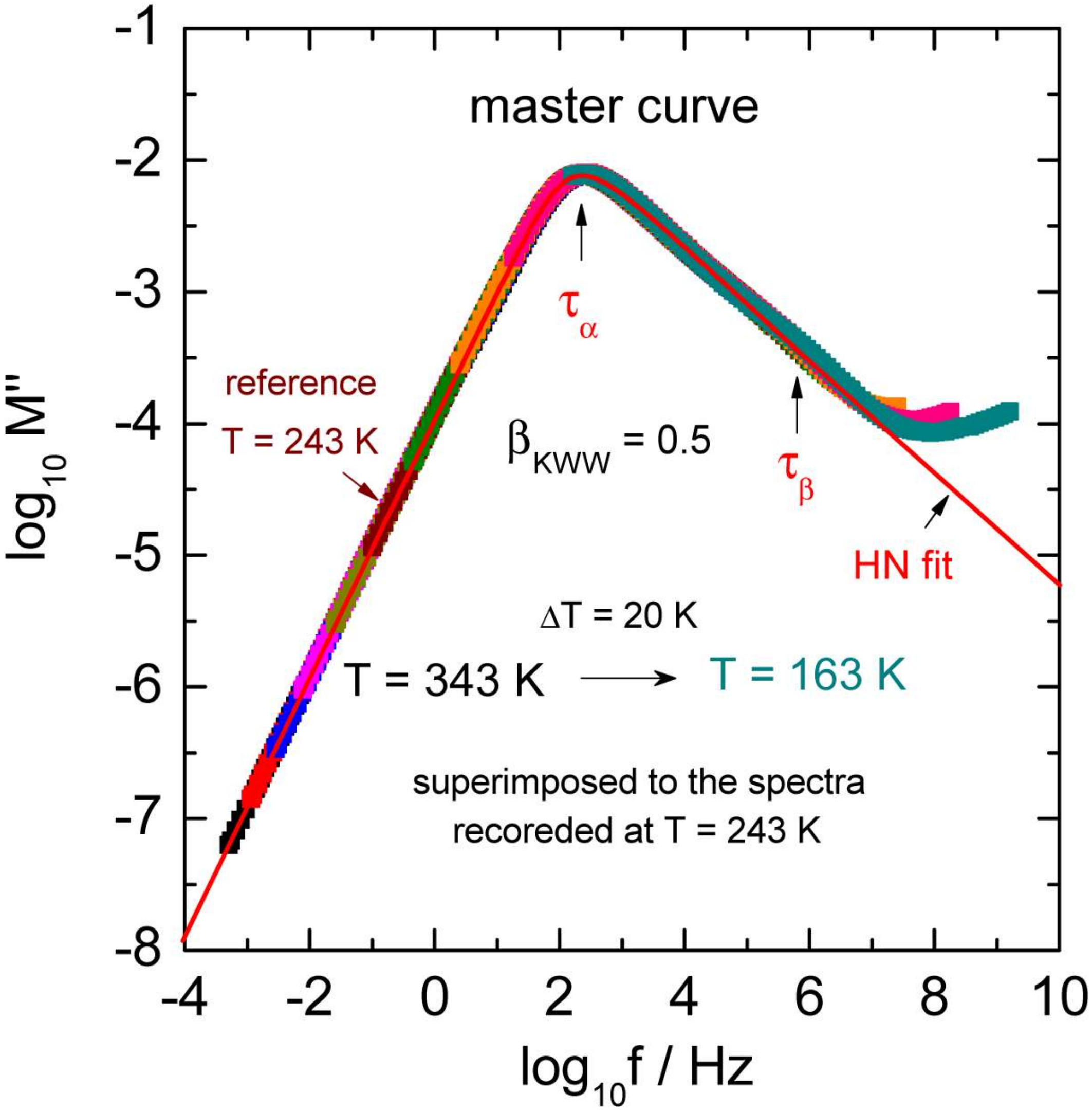

**Figure.6**

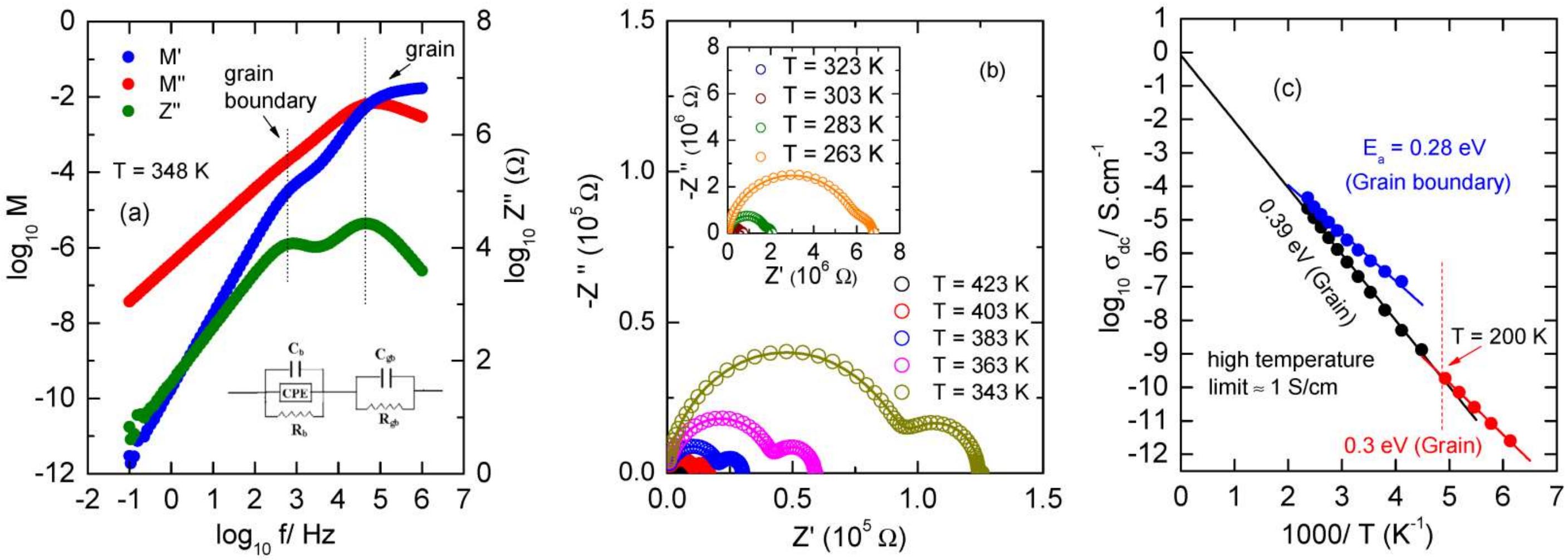

**Figure.7**

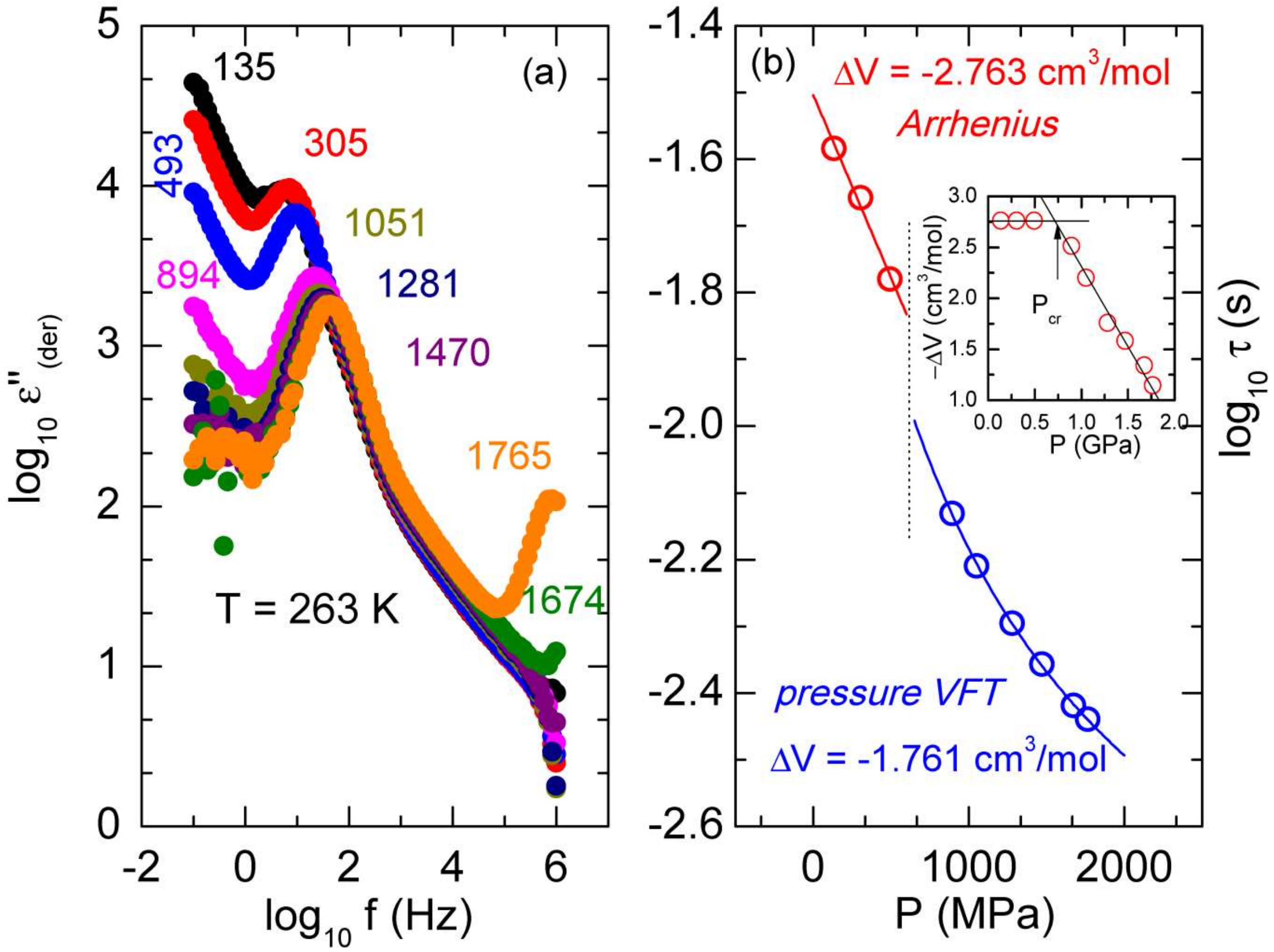

**Figure.8**

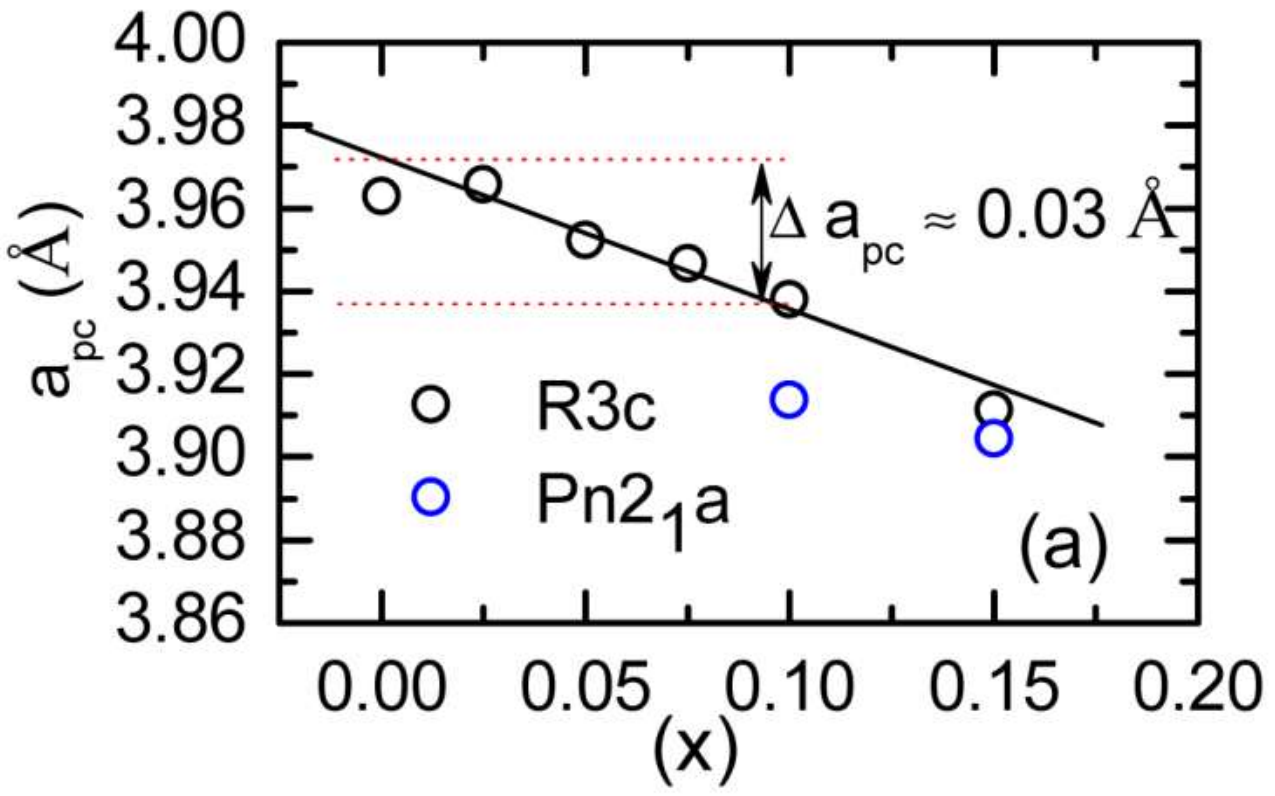

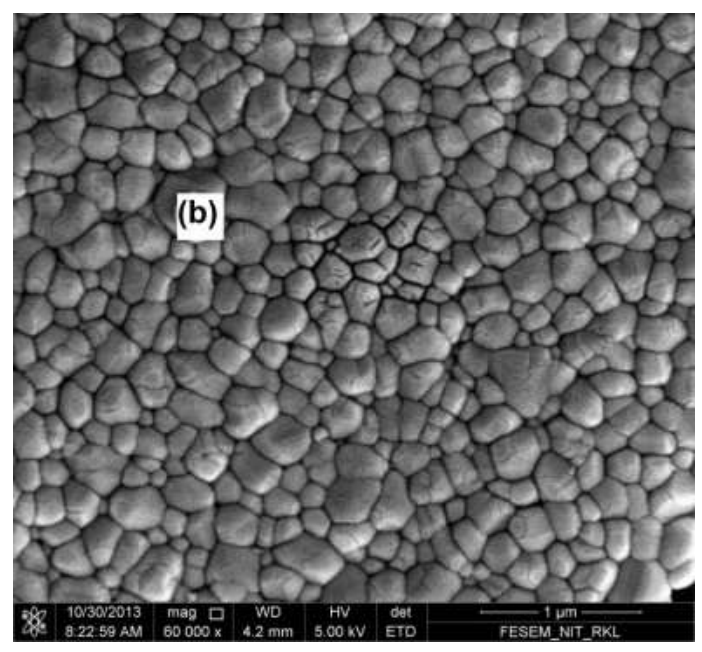

**Figure. S1**

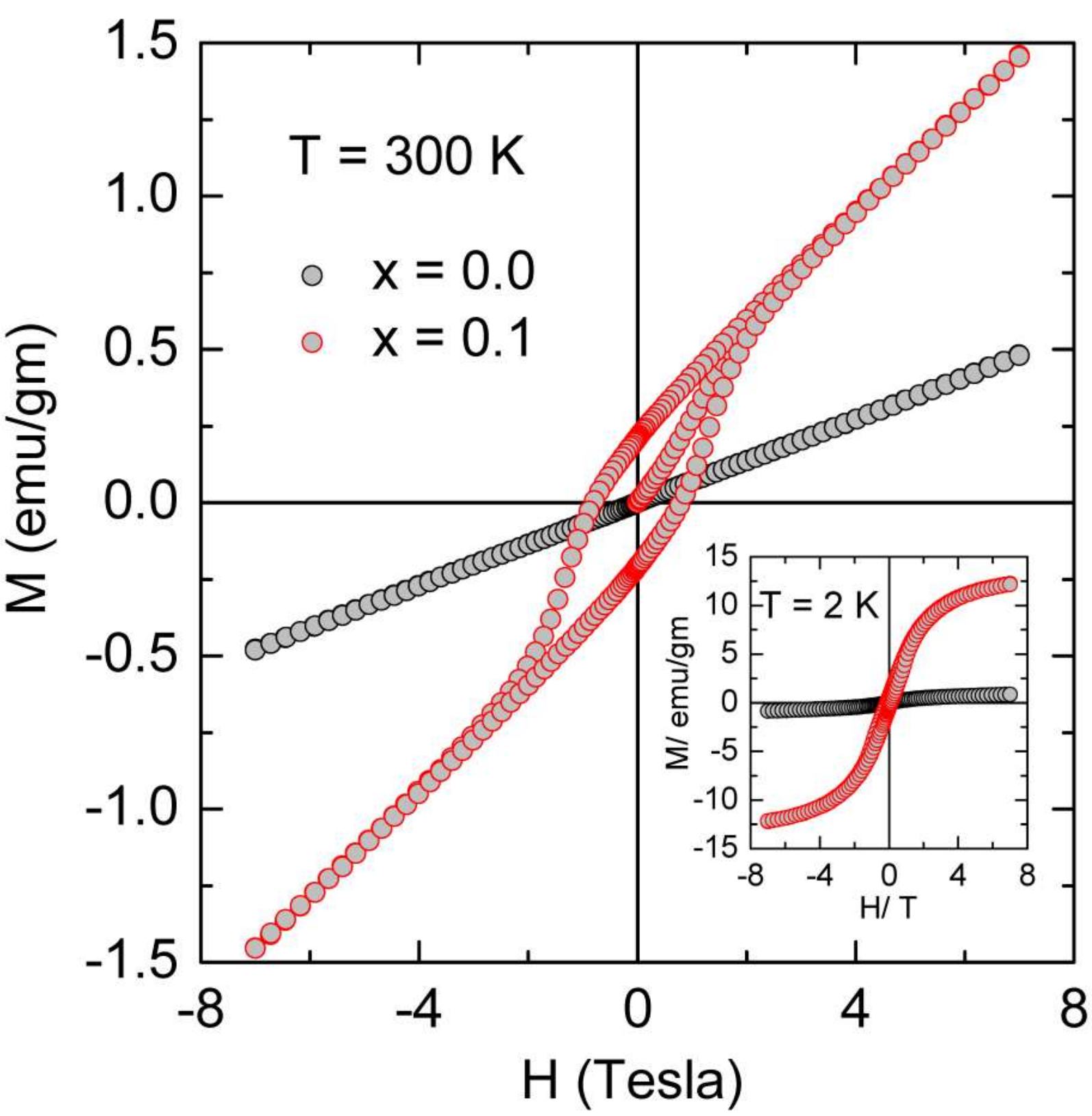

**Figure. S2**